\documentclass[reprint,aps,onecolumn,prd,noshowpacs,floatfix,nofootinbib,citeautoscript,preprintnumbers,12pt]{revtex4-1}

\usepackage{mathptmx}
\usepackage[english]{babel}
\usepackage{amssymb}
\usepackage{amsfonts}
\usepackage{amsmath}
\usepackage{eucal}
\usepackage{graphicx,color}
\usepackage{epsfig}
\usepackage[pdftex, bookmarks=true,colorlinks,linkcolor=red,urlcolor=blue,citecolor=blue]{hyperref}
\def\arXiv#1{\href{http://arxiv.org/abs/#1}{arXiv:#1}}
\def\arXiv#1#2{\href{http://arxiv.org/abs/#1}{arXiv:#1}}

\def\doi#1#2{\href{http://doi.org/#1}{#2}}

\def\be{\begin{eqnarray}}
\def\ee{\end{eqnarray}}
\def\bea{\begin{eqnarray}}
\def\eea{\end{eqnarray}}
\newcommand{\beann}{\begin{eqnarray*}}  \newcommand{\eeann}{\end{eqnarray*}}
\newcommand{\bfig}{\begin{figure}} \newcommand{\efig}{\end{figure}}
\newcommand{\ba}{\begin{array}} \newcommand{\ea}{\end{array}}
\newcommand{\bcen}{\begin{center}} \newcommand{\ecen}{\end{center}}
\newcommand{\btab}{\begin{tabular}} \newcommand{\etab}{\end{tabular}}
\newcommand{\nn}{\nonumber}
\newtheorem{Proposition}{Proposition}[section]

\newtheorem{Theorem}{Theorem}[section]
\newtheorem{Lemma}{Lemma}[section]
\newtheorem{Corrolary}{Corrolary}[section]

\newcommand{\bp}{\begin{Proposition}}	\newcommand{\ep}{\end{Proposition}}
\newcommand{\bt}{\begin{Theorem}}	\newcommand{\et}{\end{Theorem}}
\newcommand{\bl}{\begin{Lemma}}		\newcommand{\el}{\end{Lemma}}
\newcommand{\bc}{\begin{Corrolary}}	\newcommand{\ec}{\end{Corrolary}}
\begin{document}

%%%%%%%%%%%%%%%%%%%%%%%%%%%%%%%%%%%%%%
%%%%%%%%%%%%% TITLEPAGE %%%%%%%%%%%%%%%%%%
%%%%%%%%%%%%%%%%%%%%%%%%%%%%%%%%%%%%%%
\title{Anomalous transport model with axial magnetic fields}

\author{Karl Landsteiner$^{\textcolor{red}{a}}$}\email{karl.landsteiner@csic.es}
\author{Yan Liu$^{\textcolor{red}{b}}$}\email{yanliu@buaa.edu.cn}
\affiliation{$^{\textcolor{red}{a}}$Instituto de F\'{\i}sica Te\'orica UAM/CSIC, C/ Nicol\'as Cabrera
13-15,\\Universidad Aut\'onoma de Madrid, Cantoblanco, 28049 Madrid, Spain}
\affiliation{ $^{\textcolor{red}{b}}$Department of Space Science, and International Research Institute \\
of Multidisciplinary Science, Beihang University,  Beijing 100191, China}

\begin{abstract}
The transport properties of massless fermions in $3+1$ spacetime dimension have been in the
focus of recent theoretical and experimental research. 
New transport properties appear 
as consequences of chiral anomalies. The most prominent is the generation of a current in a magnetic
field, the so-called chiral magnetic effect leading to an enhancement of the electric conductivity (negative magnetoresistivity). We study the analogous effect for axial magnetic fields that couple
with opposite signs to fermions of different chirality. We emphasize local charge conservation
and study the induced magneto-conductivities proportional to an electric field and a gradient in temperature.  \
We find that the magnetoconductivity is enhanced whereas the magneto-thermoelectric conductivity is diminished. 
As a side result we interpret an anomalous contribution to the entropy current as a generalized thermal Hall effect.
\end{abstract}

\preprint{IFT-UAM/CSIC-17-020}
\maketitle

%
%%%%%%%%%%%%%%%%%%%%%%%%%%%%%%%%%%%%%%
%%%%%%%%%%%% INTRODUCTION %%%%%%%%%%%%
%%%%%%%%%%%%%%%%%%%%%%%%%%%%%%%%%%%%%%
%\section{\label{sec:intro}Introduction.}

\section{Introduction and Motivation}

Chiral anomalies \cite{Bertlmann:1996xk,Fujikawa:2004cx} and the specific transport phenomena induced by them such as the chiral magnetic and the chiral vortical effects have been extensively discussed in the recent years (see \cite{Kharzeev:2013ffa,Landsteiner:2016led} for reviews). 

In a theory of massless Dirac fermions the vector current $J^\mu = \bar \Psi \gamma^\mu \Psi$ and
axial current $J^\mu_5 = \bar \Psi \gamma_5 \gamma^\mu \Psi$ can be defined.
In such a theory the chiral magnetic effect (CME) describes the generation of an electric current in a magnetic field in 
the presence of
an axial chemical potential
\begin{equation}\label{eq:cmecov}
\vec J = \frac{\mu_5}{2\pi^2} \vec B\,,
\end{equation}
where $\mu_5$ is the axial chemical potential conjugate to the axial charge operator $Q_5 = \int d^3x \bar \Psi \gamma_5 \gamma^0 \Psi$. 

This formula has to be interpreted with care. At first sight it predicts the generation of a current in
equilibrium. It has been pointed out however that such an equilibrium current is forbidden by the so-called Bloch theorem.
In relation to the CME this theorem has first been invoked in a condensed matter context in \cite{vazifehfranz}. A recent discussion of the Bloch theorem has been given in \cite{naoki}. The theorem can be formulated as
\begin{equation}\label{ref:Blochtheorem}
\int d^3x\, \vec{J}(x) =0\,,
\end{equation}
in thermal equilibrium. Seemingly this is violated by eq. (\ref{eq:cmecov}) for a homogeneous magnetic field.
The important point emphasized in \cite{naoki} is that the Bloch theorem is valid only for exactly conserved currents. 
This allows to resolve the tension between eq. (\ref{eq:cmecov}) and the Bloch theorem.
More precisely eq. (\ref{eq:cmecov}) holds only for the so-called covariant version of the current. This covariant
current is not a truly conserved current but rather has the anomaly
\begin{equation}\label{eq:covanom}
\partial_\mu J^\mu = \frac{1}{8\pi^2} \epsilon^{\mu\nu\rho\lambda} F_{\mu\nu} F^5_{\rho\lambda}\,,
\end{equation}
where one also introduces a axial field $A_\mu^5$ as source for insertions of the axial current $J^\mu_5$. 
Similarly the covariant version of the axial anomaly is
\begin{equation}
\partial_\mu J_5^\mu = \frac{1}{16\pi^2} \epsilon^{\mu\nu\rho\lambda}\left( F_{\mu\nu} F_{\rho\lambda} +  F^5_{\mu\nu} F^5_{\rho\lambda} \right) \,.
\end{equation}
In quantum field theory the currents  are composite
operators and  have to be regularized. This regularization introduces certain ambiguities that have
to be fixed by demanding certain classical properties of the currents to hold on the quantum level. One way to fix these
ambiguities is to define $J^\mu$ and $J^\mu_5$ to be invariant objects under both vector- and axial-type gauge transformations \cite{Bardeen}. The disadvantage of this definition is that it does not result in a conserved vector like current
but rather leads to the anomaly eq. (\ref{eq:covanom}).
On the other hand one can insist on the vector like current to be exactly conserved $\partial_\mu \mathcal{J}^\mu =0$.
The relation between the two definitions of currents is
\begin{equation}\label{eq:defconscur}
\mathcal{J}^\mu = J^\mu - \frac{1}{4\pi^2} \epsilon^{\mu\nu\rho\lambda} A^5_\nu F_{\rho\lambda}\,.
\end{equation}
Due the the axial anomaly the axial vector $J^\mu_5$ is never conserved and therefore its source $A_\mu^5$ can
not be interpreted as a true gauge field. Therefore the Chern-Simons current in (\ref{eq:defconscur}) is a physical
current in a completely analogous way as the the Chern-Simons current appearing in the quantum Hall effect. 
This resolves the tension between the chiral magnetic effect and the Bloch theorem in the following manner. Thermal equilibrium is defined by the grand canonical ensemble with density matrix $\exp(-(H - \mu_5 Q_5)/T)$. This is equivalent 
to considering the theory in the background of a temporal component of the axial field $A^5_0 = \mu_5$. Now the chiral magnetic
effect in the exactly conserved current $\mathcal{J}^\mu$ takes the form \cite{Gynther:2010ed}
\begin{equation}\label{eq:cmeconscur}
\vec{\mathcal{J}} = \frac{\mu_5}{2\pi^2} \vec B - \frac{A_0^5}{2\pi^2} \vec B\,,
\end{equation} 
where the second term stems from the Chern-Simons current in eq. (\ref{eq:defconscur}). Since in strict equilibrium
$A_0^5 = \mu_5$ this shows that the chiral magnetic effect for the conserved current (\ref{eq:defconscur}) vanishes as demanded by the Bloch theorem. The importance of defining the coserved current has also been discussed in chiral kinetic theory in \cite{Gorbar:2016ygi}.

On the other hand the closely related chiral separation effect 
\begin{equation}
\vec{J}_5 = \frac{\mu}{2\pi^2} \vec B\,,
\end{equation}
does not suffer any such correction. Since the axial current is always affected by an anomaly there is no contradiction
to the Bloch theorem as pointed out in \cite{naoki}.

There is however a third related effect if one allows for axial magnetic fields,  
$\vec{B}_5 = \vec \nabla \times \vec A_5$. This is a magnetic field that couples with opposite signs
to fermions of different chirality. The axial magnetic effect takes the form
\begin{equation}\label{eq:ame}
\vec{J} = \vec{\mathcal{J}} = \frac{\mu}{2\pi^2} \vec{B}_5\,.
\end{equation}
 Formally it describes the generation of a vector-like current in the background of an axial magnetic field at finite
(vector-like) chemical potential. Note that the formula holds for both the covariant and the conserved form of the 
currents. Therefore this formula seems to be in much greater tension with the Bloch theorem than the chiral magnetic effect. 
One might dismiss this tension on the grounds that so far at a fundamental level no axial fields seem to exist in nature. However, it has been argued that such fields can appear in the effective description of the electronics of advanced materials, the so-called Weyl semimetals \cite{Cortijo:2016yph,  Pikulin:2016wfj, Grushin, Cortijo:2016wnf}.  A low energy field theoretical description of the electronics of these materials
given by the Dirac equation
\begin{equation}\label{eq:effectiveDirac}
\gamma^\mu( i D_\mu + b_\mu \gamma_5) \Psi =0\,.
\end{equation}
Here $D_\mu$ is the usual covariant derivative and the parameter $b^\mu$ enters just like the field $A^5_\mu$ coupling to  the axial current. It has been argued that straining such materials can lead to spatial variation of the
parameter $b^\mu$ and in consequence to the appearance of effective axial magnetic fields in eq. (\ref{eq:effectiveDirac}).
The reason why there is no contradiction to the Bloch theorem in this case is as follows. The parameter $b^\mu$ exists
only within the material and necessarily vanishes outside. If for definiteness we assume the axial magnetic field
to be directed along the $z$ direction and we compute the total axial flux at through a surface $\Omega$
at some fixed $z=z_0$
\begin{equation} \label{eq:phi5}
\Phi_5 = \int_\Omega dx dy B^5_z(x,y,z_0) = \int_{\partial \Omega} d\vec S \cdot \vec b =0\,,
\end{equation}
since one can always take the boundary of the surface to lie entirely outside the material where $\vec{b}=0$.
Therefore the axial analogue of the chiral magnetic effect (\ref{eq:ame}) can not induce a net current and
this resolves the tension with the Bloch theorem since no net current can be generated \cite{HosurQi,Landsteiner:2013sja}.

We will take these considerations as motivation to study electro- and thermo-magnetotransport in the background of
axial magnetic fields under the assumption that the Bloch theorem is implemented by 
a vanishing net axial magnetic flux (\ref{eq:phi5}). This implies that the net equilibrium electric current vanishes but as
we will see upon applying an electric field (or equivalently a gradient in chemical potential) and a temperature
gradient leads to anomaly induced net contributions to the currents.

\vspace{.2cm}
\section{Anomalous transport}
\label{sec3}

We study a simple of model of anomalous transport with coupled energy and charge transport. 
This means that in contrast to a full hydrodynamic model we assume that no significant collective flow
parametrized by a flow velocity develops.\footnote{This does not mean that the velocity or the variation of the velocity is zero, just that it cannot be determined by the conserved equations. Our transport model can not be obtained from hydrodynamics by setting the flow velocities to zero. Hydrodynamic flow (i.e. non vanishing velocity) appears 
already at zero order in derivatives and this imposes constraints on the first
order transport coefficients that can appear in the constitutive relations \cite{Kovtun:2012rj}. Since for strong momentum relaxation flow is absent such relations are not present.
This model has similarity to the treatment in the theory for incoherent metal in 2+1D \cite{Hartnoll:2014lpa}.} Not only is this a simpler model allowing to study the effects
of anomalies on transport it might also be more directly relevant to systems
where elastic scattering on impurities impedes the build up of collective flow.  

We develop now a formal transport model based on the anomalous continuity equations 
\begin{align}\label{eq:conservenergy}
\dot\epsilon + \vec \nabla\cdot \vec J_\epsilon &= \vec E \cdot \vec J \,,\\ 
\label{eq:conservecharge}
\dot \rho + \vec \nabla \cdot \vec J &= c \vec E\cdot \vec B\,,
\end{align}
where $\epsilon$ is the energy density and $\vec J_\epsilon$ is the energy current.
Charge conservation is affected by an anomaly with anomaly coefficient $c$. 
The right hand side of equation (\ref{eq:conservenergy}) quantifies the energy injected into the system
by an electric field (Joule heating) whereas (\ref{eq:conservecharge}) describes the (covariant) anomaly.
So far this is not specific to axial magnetic fields but rather relies only on the presence of an
anomaly in the current $J^\mu = (\rho, \vec J)$.

To discuss transport we write down constitutive relations for $\vec J_\epsilon$,  $\vec J$ and take as thermodynamic 
forces the gradients in the
thermodynamic potentials and external electric and magnetic fields, 
\begin{align}\label{eq:constitutive}
\left(
\begin{array}{c}
\vec J_\epsilon\\
\vec J\\
\end{array}
\right) =
L \cdot
\left(
\begin{array}{c}
\vec \nabla \left(\frac 1 T\right)\\
\frac{\vec E}{T} - \vec\nabla \left( \frac{\mu}{T} \right) \\
\end{array} \right) +
\left(
\begin{array}{c}
\hat \sigma_B\\
\sigma_B \\
\end{array} \right) \vec B\,.
\end{align}

The matrix $L$ encodes response due to gradients in chemical potential and temperature. $\{\hat \sigma_B, \sigma_B\}$ 
describe response due to the magnetic field. In principle we could also allow an independent response due to the electric field. In our ansatz we have thus anticipated that positivity of entropy production is not compatible with
such additional terms in the constitutive relations.

The transport coefficients are constrained by the second law of thermodynamics. Using the thermodynamic 
relation $T ds = d\epsilon + \mu d\rho$ as guideline we define the entropy current \cite{Son:2009tf} as
\begin{equation}\label{eq:entropycurrent}
\vec J_s = \frac 1 T \vec{J}_\epsilon - \frac \mu T \vec J + \eta_B \vec B\,.
\end{equation}
Up to the terms depending on the magnetic field this is the standard ansatz for
coupled energy and charge transport \cite{LeBellac}. 

Following \cite{Son:2009tf,Neiman:2010zi} we impose the local form of the second law thermodynamics
\begin{equation}\label{eq:2nd}
\dot s + \vec \nabla \cdot \vec J_s \geq 0\,.
\end{equation}
Using $T \dot s =  \dot \epsilon + \mu \dot \rho$ this leads to
\begin{equation}
\frac{1}{T} \left( \frac{\partial\epsilon}{\partial t} + \vec \nabla \cdot\vec J \right) - \frac{\mu}{T} \left(
\frac{\partial \rho}{\partial t} + \vec \nabla \cdot \vec J \right)  + \vec \nabla \left(\frac{1}{T}\right) \cdot\vec J_\epsilon
 - \vec \nabla \left( \frac{\mu}{T}\right) \cdot \vec J + \vec \nabla \eta_B \cdot \vec B + \eta_B \nabla\cdot\vec B \geq 0\,.
\end{equation}
We assume absence of magnetic monopoles and thus the last term vanishes.  
Using the constitutive relations 
we find the constraints $det(L)\geq 0$ and $L_{11}\geq 0$ and $L_{22}\geq 0$. 
Positivity of the entropy production also assures that the
electric field does not give rise to additional response not already contained in $L$.
Entropy is produced only by the symmetric part of the matrix $L$.
%and Onsager's reciprocity relations guarantee that $L$ is symmetric in the absence of time reversal breaking. 
For the magnetic conductivities one finds a set of one algebraic and two differential equations
\begin{align}
\sigma_B - c \mu &= 0\,,\\
\sigma_B + \frac{\partial \eta_B}{\partial \gamma_\rho} &=0\,,\\
\hat\sigma_B + \frac{\partial \eta_B}{\partial \gamma_\epsilon} &=0\,,
\end{align}
with $\gamma_\epsilon = 1/T$ and $\gamma_\rho = - \mu/T$. 
These equations are the
coefficients of the terms $(\vec E\cdot \vec B)$, $(\vec \nabla \gamma_\epsilon \cdot \vec B_5)$ and 
$(\vec \nabla \gamma_\rho \cdot \vec B)$.
These terms can be either positive or negative and therefore their coefficients must vanish to guarantee the local form of the second law of thermodynamics. 
Since there is no further dimensionful parameter $\eta_B$ must also fulfill $\gamma_\epsilon\partial \eta_B / \partial \gamma_\epsilon = - \eta_B$ as it has to have dimension one, where in our conventions $(\mu, T)$ have dimension one.
The magnetic conductivities are almost completely determined 
\begin{align}\label{eq:cmcs}
\sigma_B =  c \mu\,,~~\hat\sigma_B = c \frac{\mu^2}{2} + c_g T^2\,,~~~ \eta_B = c \frac{\mu^2}{2 T} + c_g T\,.
\end{align} 
Up to ambiguities arising due to frame choice these are basically the same results as in hydrodynamics \cite{Son:2009tf, Neiman:2010zi, Stephanov:2015roa}.

The priori undetermined integration constant $c_g$ is related to (mixed) gravitational anomalies \cite{Landsteiner:2011cp, Landsteiner:2011iq,Golkar:2012kb,Jensen:2012kj,Golkar:2015oxw, Jensen:2013rga, Chowdhury:2016cmh, Glorioso:2017lcn, Basar:2013qia}. 
%In kinetic theory the relation to (mixed) gravitational anomalies was shown in \cite{Basar:2013qia}.
In holography it was also shown recently that the relation to the (mixed) gravitational anomaly is not modified by
momentum relaxation in \cite{Copetti:2017ywz}. The intuition that dissipationless 
transport should not be affected by momentum relaxation together with the
results of \cite{Copetti:2017ywz} and \cite{Stephanov:2015roa} (the case of weak momentum relaxation) we take 
as evidence that $c_g \neq 0$  also in the case of strong momentum relaxation and
that it is related to the presence of (possibly global) gravitational 
anomalies. 
For theories containing only spin $1/2$ particles and holographic theories
this relation is $c_g = 32 \pi^2 \lambda$ where $\lambda$ is the coefficient of the gravitational contribution to
the anomaly $\partial_\mu J^\mu = \lambda \epsilon^{\mu\nu\rho\lambda} R^\alpha\,_{\beta\mu\nu} R^\beta\,_{\alpha\mu\nu} $.
A single Weyl fermion has $\lambda=\pm\frac{1}{768\pi^2}$ and $c_g=\pm 1/24$ with the sign depending on the chirality. 
In the following assume $c_g\neq 0$ to be related to the mixed gravitational anomaly as in the case without momentum relaxation and study its implications for thermo-electric transport in axial magnetic fields.

Using the results for the anomalous transport coefficients $\sigma_B$, $\hat \sigma_B$ and $\eta_B$ the 
entropy current can be written as
\begin{equation}\label{eq:entropycur}
\vec J_s = (1/T, -\mu/T)\cdot L \cdot \left(\begin{array}{c}\vec\nabla (1/T)\\
\frac{\vec{E}}{T}-\vec \nabla\big( \frac{\mu}{T}\big) \end{array}\right) +  2 c_g T \vec B\,.
\end{equation}
Naively one might have expected that the anomalous transport does not contribute to the entropy current. 
It turns out however that the temperature dependence encoding the gravitational anomalies does contribute
to entropy current. This has been previously observed in \cite{Chapman:2012my,Stephanov:2015roa}. 

The previous considerations are general and assume only the presence of an anomaly in the charge current.
We can now specialize to the case of the axial magnetic field. In this case the charge conservation takes the form
\begin{equation}
\dot \rho + \vec\nabla \cdot \vec J = \frac{N_f}{2 \pi^2} (\vec E \cdot \vec B_5 + \vec{E}_5 \cdot \vec{B} )\,,
\end{equation}
where $N_f$ is the number of Dirac fermions. Rather than an anomaly in this case the right hand side should
be interpreted as the divergence of the Chern-Simons current in eq. (\ref{eq:defconscur}). 
Since we are mostly concerned with the effects of axial magnetic fields we will set $\vec{E}_5=\vec B =0$ in the following.
In this case the conservation equations are precisely as in the general case before and we can take over the previous results by simply replacing $\vec B$ with $\vec B_5$ and setting $c=\frac{N_f}{2\pi^2}$ and $c_g=N_f/12$.

Now we want to give an interpretation for the anomalous contribution to the entropy current in eq. (\ref{eq:entropycur}).
We consider an axial magnetic field configuration of the form
\begin{equation}
\vec B_5(x) = \hat e_z\bar \Phi_5 \big[ \delta(x) - \delta(x-L)\big].
\end{equation}
According to our assumption of compatibility with Bloch's theorem the total axial magnetic flux along the $z$ direction
vanishes but the regions of positive and negative fluxes are well separated which for simplicity we model by delta-functions
distribution localized in $x=0$ and $x=L$ but spread out in the $y$ direction. 
The first thing to notice is that
according to (\ref{eq:entropycur}) there is an anomalous entropy current localized at the locations of axial magnetic flux.
If there is a temperature gradient such that $T(x=0)=T+ \delta T$ and $T(x=L)=T$ a net entropy current flows along the $z$-direction 
\begin{equation}
\delta \vec I_s = \int dx  \vec J_s  = 2 c_g (\delta T)  \Phi_5 \hat e_z\,.
\end{equation}
This current flows in a direction orthogonal to the temperature gradient. 
Heat is not a thermodynamic state variable still it can be defined as $\delta Q = T dS$ and in an analogous way
we can define a heat current as $\delta \vec{I}_Q = T \delta \vec I_s$. This leads to the net heat current
\begin{equation}
\delta \vec{I}_Q = 2 c_g T \delta T \bar \Phi_5 \hat e_z\,.
\end{equation} 
We interpret this as anomalous thermal Hall effect. In this way 
the anomalous contribution to the entropy current in (\ref{eq:entropycur})  can be understood as a generalization of the anomalous thermal Hall effect.
Previous discussions of the relation between thermal Hall effect and gravitational 
anomalies are \cite{Nakai:2016lle, Stone:2012ud}. Let us also note that the very concept of heat current can be questioned
on the grounds that heat is not a state variable \cite{Romer}. In the context of anomalous transport there certainly 
arises the question if in the common definition of heat current $\vec J_Q = \vec J_\epsilon - \mu \vec J$ the current
$\vec J$ should be taken to be the covariant or the conserved current. Defining the heat current as 
$\delta \vec J_Q = T \delta \vec J_s$ resolves this issue. 

\subsection{Induced conductivities}

Let us now come to the main subject: the linear response of this system to a temperature gradient and an external electric field both aligned with the axial magnetic field. 
The continuity equations (\ref{eq:conservenergy}), (\ref{eq:conservecharge}) together with the constitutive relations
(\ref{eq:constitutive}) form a dynamical system that allows to compute current and charge distributions given some initial
and boundary conditions. The effective response to an applied electric field and a temperature gradient can be computed 
by solving these equations.

Before studying the axial magnetic field case of interest it is worth to briefly recall how the chiral magnetic effect
leads to negative magneto-resisitivity \cite{Nielsen:1983rb, Son:2012bg,Li:2014bha}. One assumes a homogeneous magnetic field and a parallel electric field.
Axial charge is not subject to an exact conservation law and thus it is natural to introduce an axial charge
relaxation time $\tau_5$. Non-conservation of axial charge is provided e.g. by a mass term in the Dirac equation or
by inter-valley scattering the context of Weyl semimetals.
The effective axial charge (non-)conservation is then
\begin{equation}\label{eq:axialdecay}
\dot \rho_5 = c \vec E \cdot \vec B - \frac{1}{\tau_5} \rho_5\,.
\end{equation}
We note that if an external electric field is absent but instead a gradient of the chemical potential is induced
the chiral magnetic current has a non-vanishing gradient $\vec \nabla\cdot \vec J = c \vec \nabla \mu \cdot \vec B$ 
which leads to en effectively equivalent equation for the time development of axial charge by replacing $\vec E \rightarrow -\vec \nabla \mu $. Axial charge is built up until a steady state is reached with $\delta\rho_5 = \tau_5 c \vec E \cdot \vec B$. The axial charge can be related to the axial chemical potential via $\chi_5 \delta\mu_5 = \delta\rho_5$ where
$\chi_5$ is the axial susceptibility. Combining Ohmic and chiral magnetic currents leads to the enhanced current
\begin{equation}
\vec J = \sigma \vec E + \tau_5\frac{c^2 (\vec E\cdot \vec B)}{\chi_5 }  \vec B\,.
\end{equation}
For infinite axial charge relaxation time the anomaly induced magnetoconductivity is formally infinite and this might
be referred to as chiral magnetic superconductivity \cite{Kharzeev:2016tvd}. In nature fermions are however massive and effective chiral
fermions in materials such as Weyl semimetals do not preserve there chirality at all energies due to the compact nature
of of the Brillouin zone.  

In the case of the axial magnetic field the role of the axial chemical potential is played by the (electric)
chemical potential $\mu$. Electric charge is an exactly conserved quantity due to electro-magnetic gauge invariance.
Therefore it is not possible to introduce a relaxation time for electric charge without violating gauge invariance.
If it were possible then to engineer homogeneous axial magnetic fields an analogous argument would lead necessarily
to infinite axial magneto-conductivity. As we have argued however in the introduction the assumption of such a homogeneous
axial magnetic field is by itself inconsistent with the Bloch theorem, which by itself is a consequence of gauge invariance \cite{naoki}.
Thus we are naturally lead to study induced electro- and thermo- axial magneto conductivity under the constraint of 
vanishing net axial magnetic flux. This makes the problem more complicated as diffusion from regions where charge
is accumulated to regions with charge outflow has to be taken into account. It is this diffusion process that
can lead to a stationary state and finite induced axial magneto-conductivities.

As external driving forces we assume a homogeneous electric field and a temperature gradient pointing in the 
$z$ direction. We also assume an axial magnetic field directed along the $z$ direction 
but inhomogeneous in the $(x,y)$ plane and with zero net flux $\Phi_5 = \int dx dy B_5 (x,y) =0$.
The dynamical variables are the chemical potential $\mu$ and the temperature $T$.
We allow the system to adjust to the external forces by developing non-trivial profiles  of chemical
potential and temperature in the $(x,y)$ plane around a constant background value. Thus our ansatz is
\begin{align}
\vec B_5&=  B_5 (x,y) \hat e_z\,, ~~~~~~~~~~ \vec E = E \hat e_z\,, \\
 \mu &= \mu_0 + \delta \mu(x,y) \,,~~~
T = T_0 + \delta T(x,y) + z\nabla_z T \,.
\end{align}
The response in $\delta \mu$ and $\delta T$ to $E$ and $\nabla T$  is now calculated in linear approximation.

Since the axial magnetic field is not uniform in the $(x,y)$ plane the system will react to the local charge inflow
induced by the anomalous Hall and axial magnetic effects by building up diffusion currents. 
Eventually a stationary state is reached. This stationary state can be obtained from the constitutive relations and the conservation equations by dropping the time derivative. 
We furthermore assume the matrix $L$ to be spatially isotropic. 
Using (\ref{eq:conservenergy}), (\ref{eq:conservecharge}) the constitutive relations (\ref{eq:constitutive}) with the 
anomalous transport coefficients (\ref{eq:cmcs}) we find that the fluctuations $\delta T$ and $\delta \mu$
have to fulfill a system of Poisson equations
\begin{equation}\label{eq:diffusion}
L\cdot Y \Delta_\perp \left(
\begin{array}{c}
\delta T(x_\perp)\\
\delta \mu (x_\perp)
\end{array}
\right) =
\left( \begin{array}{cc}
- 2 c_g T_0 & c \mu_0\\
0& c
\end{array}\right) \cdot
\left(
\begin{array}{c}
\nabla_z T\\
E
\end{array}
\right) B_5(x_\perp)\,.
\end{equation}
Here $\Delta_\perp$ is the two dimensional Laplace operator ($\Delta_\perp=\partial_x^2+\partial_y^2$) and $Y=\frac{1}{T_0^2}\left(\begin{array}{cc}
-1&0\\
\mu_0&-T_0
\end{array}\right)$ 
is the transformation matrix relating the the thermodynamic forces $\delta(1/T)$ and $\delta(-\mu/T)$ to the
fluctuations $\delta T$, $\delta \mu$. Once the fluctuations are determined they can be plugged into the {\em anomalous
part} of constitutive relations (\ref{eq:constitutive}) to find the anomaly induced contribution to the currents
\begin{equation}
\label{eq:tran-def}
\left(
\begin{array}{c}
J^z_\epsilon \\
J^z
\end{array}
\right)_\mathrm{anom} =
- B_5(x_\perp) u(x_\perp) \Sigma \cdot \left(
\begin{array}{c}
\nabla^z (\frac 1 T) \\
\frac{E}{T} - \nabla^z (\frac{\mu}{T})
\end{array}
\right) \,,
\end{equation}
with the conductivity matrix
\begin{equation}\label{eq:cond}
\Sigma = \left( \begin{array}{cc}
2 c_g T_0 & c \mu_0\\
0& c
\end{array}\right) \cdot (L\cdot Y)^{-1} \cdot \left( \begin{array}{cc}
-2 c_g T_0 & c \mu_0\\
0& c
\end{array}\right)\cdot
\left( \begin{array}{cc}
T_0^2 & 0\\
-T_0\mu&-T_0
\end{array}\right)
\end{equation}
and the solution to the Poisson equation $\Delta_\perp u(x_\perp)=B_5(x_\perp)$, i.e.
\begin{equation}\label{eq:solu}
u (x_\perp) = \int dx'_\perp G(x_\perp - x'_\perp) B_5(x'_\perp)\,.
\end{equation}

We have written the induced conductivity matrix as acting on the naturally defined thermodynamic forces. This
has the advantage that the Onsager reciprocity relations are automatically satisfied, i.e. $\Sigma$ is symmetric, 
\begin{align}
\label{eq:sigma11} \Sigma_{11} =& \frac{1}{\mathrm{det}(L)}T^2\Big( (L_{22}(2 c_gT^2+c\mu^2)^2+ c^2\mu^2 L_{11} )-\nn\\
&~~~
-2 c \mu (2 c_g T^2+c\mu^2) L_{12} \Big)\,,\\
%\Sigma_{11} =& \frac{1}{\mathrm{det}(L)}T^2 \Big(4 c_g^2 T^4L_{22} + 4 c c_g T^2\mu (-L_{12}+L_{22}\mu) +\nn\\
%\label{eq:sigma11}&~~~
%c^2\mu^2(L_{11}-2 \mu L_{12}+\mu^2 L_{22}  ) \Big)\,,\\
\label{eq:sigma22}
\Sigma_{22} =& \frac{1}{\mathrm{det}(L)} c^2 T^2 \Big(L_{11}-2\mu L_{12}+\mu^2 L_{22}\Big)\,,\\
\label{eq:sigma12} 
\Sigma_{12} =& \,\Sigma_{21} =\frac{1}{\mathrm{det}(L)} c T^2\Big(2 c_g T^2(\mu L_{22}-L_{12}) + \nn\\
&~~~
c \mu (L_{11}-2 \mu L_{12}+\mu^2 L_{22})
\Big)\,.
\end{align}

Using $a+b\geq 2\sqrt{ab}$ for $a,b\geq 0$ and the fact $L_{11}\geq 0, L_{22}\geq 0, \det(L)\geq 0$ one shows that 
$\Sigma_{11}$ and $\Sigma_{22}$ are positive. Furthermore the total current is proportional 
to the expression
\begin{equation}
- \int d^2x_\perp d^2x'_\perp B_5(x_\perp) u(x'_\perp) = \int \frac{d^2q}{(2\pi)^2}
\frac{\tilde B_5(-q) \tilde B_5(q) }{q^2}
\end{equation}
which for a real function $B_5(x_\perp)$ is positive definite. Thus the response matrix
in the net current described by (\ref{eq:cond}) has the same properties as $L$ since its 
determinant is also positive
\begin{equation}
\mathrm{det}(\Sigma) = \frac{1}{\mathrm{det}(L)} 4 c^2 c_g^2 T^8\,.
\end{equation}

The electric and thermoelectric conductivity is defined as $ \vec J = \sigma \vec E - \alpha \vec\nabla T $. The thermoelectric conductivity $\alpha$
is non-vanishing only because of the contribution of the mixed axial-gravitational anomaly, 
\begin{align}
\sigma &= \sigma_0 - u B_5 \frac{\Sigma_{22}}{T} \,, \\ 
\alpha &= \alpha_0\Big( 1 + u B_5\frac{1}{\mathrm{det}(L)} 2 c c_g T^4 \Big) \,,
\end{align}
with $\sigma_0 = L_{22}/T$ and $\alpha_0 = (L_{12}- \mu L_{22})/T^2$. 
Measuring therefore the total current induced by a temperature gradient in the background
of an axial magnetic field is an experimental signature of the mixed axial-gravitational
anomaly. Note however that the electric conductivity is enhanced whereas the thermoelectric
conductivity gets diminished. This is in contrast to the anomaly induced thermoelectric conductivity
in a usual magnetic field \cite{Lundgren:2014hra, spivak, Lucas:2016omy, Gooth:2017mbd}.

\subsection{Example}

Finally we would like to discuss a simple
example demonstrating the finiteness of the total induced current. 
We assume the periodic axial magnetic field configuration
%\comment{dimensional analysis, `L' should appear?}
%\begin{equation}
%\vec B_5(x) = \hat e_z\bar B_5\comment{L} \big[ \delta(x) - \delta(x-L)\big].
%\end{equation}
%
%The solution to the Poisson equation is
%\begin{equation}
%u(x) = \bar B_5 \comment{L}  \big[x \Theta(x) - (x-L)\Theta(x-L) - \frac{L}{2}\big]\,,
%\end{equation}
%where the integration constants are fixed by demanding regularity at large $x$ and that $\int_0^L u(x) dx=0$. The latter guarantees that no net chemical potential is generated. 
%The product
%\begin{equation}
%-B_{5,z}(x) u(x) =  \bar B_5^2 \frac{L\comment{L^3}}{2}\big[\delta(x) + \delta(x-L)\big]\,,
%\end{equation}
%is positive as expected. The total induced currents are therefore proportional to
%\begin{equation}
%I_\mathrm{net} \propto -\int_0^L dx B_5(x) u(x) = \bar B_5^2 L\comment{L^3}\,.
%\end{equation}
%As a second example we consider a periodic axial magnetic field 
of the form
\begin{equation}
\vec B_5 = \hat e_z \bar B_5 \sin(2\pi x/L)\,.
\end{equation}
Integrating over a period the net flux vanishes. The solution to the Poisson equation is
now 
\begin{equation}
u = - \bar B_5 \frac{L^2}{4\pi^2} \sin(2\pi x/L)\,.
\end{equation}
As boundary conditions we have imposed that no chemical potential is induced over one period.
Now the net current density over one period of oscillation is proportional to
\begin{equation}
-\frac{1}{L}\int_0^L  dx u(x) B_{5,z}(x) = (\bar B_5)^2 \frac{L^2}{8 \pi^2}\,.
\end{equation}
The gradient in this field configuration is proportional to the inverse of the period $L$. 
The current density is therefore inversely proportional to the square of the field gradient as expected and diverges 
in the limit of homogeneous field $L\rightarrow \infty$. In this limit diffusion is not effective and since there is no relaxation of the exactly conserved electric charge we end up again with infinite conductivities. 

\vspace{.2cm}
\section{Discussion}
\label{sec4}

We have developed a simple model of coupled energy and charge transport for chiral fermions 
in the background of axial magnetic fields. 
Our study was motivated by considerations based on compatibility with the Bloch theorem that forbids
net currents in thermal equilibrium. In order to circumvent this we assumed axial magnetic field configurations
with vanishing net flux such that in equilibrium the integrated total current vanishes. 
The anomalous transport model was constructed demanding a positive definite entropy production.
Even without assuming full hydrodynamics, i.e. assuming that no significant collective flow can develop we found
that anomalies induce chiral magnetic charge and energy currents. The form of the chiral magnetic transport
coefficients contain a priori undetermined integration constant depending on the temperature which can be
related to the presence mixed gauge perturbative and global gravitational anomalies. 
As previously observed in \cite{Chapman:2012my, Stephanov:2015roa} the entropy current contains somewhat unexpectedly an anomalous term. We gave
a physical interpretation relating it to a generalized form of the thermal Hall effect, i.e. the generation of
a heat current perpendicular to a temperature gradient. 

Then we studied electro- and thermo-magneto conductivities. We found that the assumption of vanishing net axial magnetic flux activates the diffusion terms in the constitutive relations leading to finite induced conductivities.  
Despite the fact there is no net magnetic flux a net electric current is induced
either by an external electric field or by a temperature gradient.  
The resulting net axial magneto conductivity is enhanced whereas the axial 
thermo-magneto conductivity is diminished and is proportional to the 
coefficient of the mixed axial-gravitational anomaly.

The  focus in the previous literature is considers the effects due to anomalies in the presence of background magnetic fields. Anomaly related enhancement of electric and thermoelectric conductivities in magnetic fields have indeed been
observed in \cite{Li:2014bha,Gooth:2017mbd}.
In contrast in this work we have concentrated on the observable effects in the presence of background axial magnetic fields.
Our study differs in two important points from previous ones \cite{Pikulin:2016wfj,Grushin} in that we take the Bloch theorem into account and also study the thermo-electric conductivity. We hope that the effects can be measured in the future and will enrich our current understanding of the role of chiral anomaly.

\subsection*{Acknowledgments}
We thank Y.-W. Sun, S. Golkar and S. Sethi for useful discussions. The research of K.L. has been supported by FPA2015-65480-P and by the Centro de 
Excelencia Severo Ochoa Programme under grant SEV-2012-0249 and SEV-2016-0597. The research of Y.L. has 
been supported by the Thousand Young Talents Program of China and grants 
ZG216S17A5 and KG12003301 
from Beihang University. %}

\end{document}